\documentclass[useAMS,usenatbib]{mn2e}

\usepackage{subfigure}
\usepackage{graphicx}
\usepackage{times}
\usepackage{tabularx}
\usepackage{natbib}
\usepackage{url} 
\usepackage{xcolor}
\usepackage{caption}
\usepackage{amsmath}
\usepackage{float}
\usepackage{multirow}
\usepackage{morefloats}

\newcommand{\enzo}{{\it {\small ENZO}}}
\newcommand{\CRaTer}{{\it {\small CRaTer}}}
\newcommand{\MUSIC}{{\it {\small MUSIC}}}

\newcommand{\Mpc}{\mathrm{Mpc}}
\newcommand{\Msun}{\mathrm{M}_{\odot}}
\newcommand{\Mth}{\mathrm{M}_{200}}

\newcommand{\kpc}{\mathrm{kpc}}

\newcommand{\ph}{\mathrm{ph}}
\newcommand{\cm}{\mathrm{cm}}

\newcommand{\sek}{\mathrm{s}}

\newcommand{\MeV}{\mathrm{MeV}}
\newcommand{\GeV}{\mathrm{GeV}}

\newcommand{\Gyr}{\mathrm{Gyr}}

\newcommand{\para}{\mathrm{para}}

\newcommand{\pre}{\mathrm{pre}}
\newcommand{\post}{\mathrm{post}}

\newcommand{\bvec}{\mathbf{B}}
\newcommand{\vvec}{\mathbf{v}}
\newcommand{\sanpedro}{\textit{San Pedro}}

\begin{document}
 \definecolor{myred}{rgb}{1,0,0} 
 \definecolor{myblue}{rgb}{0,0,1}

 \title[$\gamma$-rays in Galaxy Clusters]{Limiting the shock acceleration of cosmic-ray protons in the ICM}
 \author[D. Wittor, ]{D. Wittor$^{1,2,3}$\thanks{%
 E-mail: denis.wittor@unibo.it}, F. Vazza$^{1,2,3}$, D. Ryu$^{4}$, H. Kang$^{5}$\\
 %EndAName
 $^{1}$Dipartimento di Fisica e Astronomia, Universita di Bologna, Via Gobetti 93/2, 40122, Bologna, Italy \\
 $^{2}$ INAF, Istituto di Radioastronomia di Bologna, via Gobetti 101, I-41029 Bologna, Italy \\
 $^{3}$ Hamburger Sternwarte, Gojenbergsweg 112, 21029 Hamburg, Germany \\
 $^{4}$ Department of Physics, School of Natural Sciences UNIST, Ulsan 44919, Korea \\
 $^{5}$ Department of Earth Sciences, Pusan National University, Busan 46241, Korea}
 \date{ Accepted 2020 April 13. Received 2020 March 30; in original form 2020 March 2.}
 \maketitle

 \begin{abstract}
  Observations of large-scale radio emissions prove the existence of shock accelerated cosmic-ray electrons in galaxy clusters, while the lack of detected $\gamma$-rays limits the acceleration of cosmic-ray protons in galaxy clusters. This challenges our understanding of how diffusive shock acceleration works. In this work, we couple the most updated recipes for shock acceleration in the intracluster medium to state-of-the-art magneto-hydrodynamical simulations of massive galaxy clusters. Furthermore, we use passive tracer particles to follow the evolution of accelerated cosmic-rays. We show that when the interplay between magnetic field topology and the feedback from accelerated cosmic rays is taken into account, the latest developments of particle acceleration theory give results which are compatible with observational constraints.
 \end{abstract}
 \label{firstpage}
 \begin{keywords}
  galaxy cluster, cosmic-ray protons, shock acceleration, $\gamma$-rays
 \end{keywords}
 \section{Introduction}\label{sec::intro}
 Some of the Universe's largest particles accelerators are found in galaxy clusters.  During the process of hierarchical structure formation, both shock waves and turbulence, that are observed by X-ray observations, form in the intracluster medium (ICM). Radio relics are large and elongated sources located at the clusters' periphery \citep[e.g.][]{2019SSRv..215...16V}. A likely explanation for relics is Diffusive Shock Acceleration \citep[DSA, e.g.][and references therin]{bykov2019rev}. Yet, several questions for the complete understanding of relics remain.  If we were to assume that DSA operates similarly for both electrons and protons, then cosmic-ray protons would be expected to fill the cluster-wide volume due to their long lifetime. However, to-date the Fermi Large Area Telescope (hereafter "Fermi") has not detected the $\gamma$-ray signal produced by inelastic collisions with the thermal protons, which limits the total energy of cosmic-ray protons to be less than a few percent of the total gas energy within clusters \citep{2014ApJ78718A,2015ApJ812159A,2016ApJ819149A}. \\
 Several works have investigated the missing $\gamma$-rays. \citet[][]{kj07} and \citet[][]{2013ApJ...764...95K} used 1D diffusion-convection equations of shocks to derive the Mach number dependent acceleration efficiencies. Yet, cosmological simulations showed that these efficiencies are too large and would produce a $\gamma$-ray signal still observable by Fermi, and that an overall efficiency of $\leq 10^{-3}$ is required to make clusters invisible in $\gamma$-rays. \citep{2016MNRAS.459...70V}. Recently, Particle-in-Cell (PIC)  simulations have shown that the shock acceleration efficiencies do not only depend on the shock strength but also on its obliquity, the angle between shock normal and the local magnetic fields \citep[][]{2014ApJ...783...91C,Guo_eta_al_2014_I,Guo_eta_al_2014_II,Kang2019electrons}. These works showed that cosmic-ray protons require a rather parallel alignment, while cosmic-ray electrons prefer a more perpendicular orientation to be efficiently accelerated. In \citet{2016Galax...4...71W,2017MNRAS.464.4448W},we found that the distribution of obliquities follows the distribution of random angles in a 3-dimensional space and, therefore, about $2/3$ of all shocks are expected to be quasi-perpendicular and only $1/3$ of all shocks tend to be quasi-parallel. While the radio emission stays unaffected by this, the $\gamma$-ray emission drops by a factor of about $\sim 3$, which cannot explain the non-detection by Fermi. \\
 More recent PIC simulations by \citet{Ha2018protons} showed that proton acceleration by low Mach number shocks in high $\beta$ plasmas (as the ICM) is quenched for shocks with Mach numbers below $\sim 2.25$. In addition, \citet{Ryu2019} included the dynamical feedback of cosmic-ray pressure on the shock and derived new acceleration efficiencies for cluster shocks. Their model predicts that the acceleration efficiencies in the Mach number regime of $2.25-5.0$ are in the range of $10^{-3}$ to $10^{-2}$. Using grid simulations, \citet{Ha2019gammas} found that these new findings produce $\gamma$-ray emission that is invisible to Fermi. Though, they do not follow the evolution of cosmic-ray protons throughout their simulation. 
 Hence in this contribution, we used methods, already presented in \citet{2016Galax...4...71W,2017MNRAS.464.4448W} but including the new constrains on the minimum Mach number \citep{Ha2018protons} and the acceleration efficiencies \citep{Ryu2019} to study their effect on the $\gamma$-ray emission. Furthermore, we applied our modelling to a new set of high-resolution magneto-hydrodynamical (MHD) simulations. This work is structured as follows: first, we introduce our simulations in Sec. \ref{sec::simu}. In Sec. \ref{sec::res}, we present our results and we conclude our work in Sec. \ref{sec::conc}.
 \section{Simulations}\label{sec::simu}
 \subsection{\enzo-simulations}
 Here, we analyzed clusters that were simulated with the MHD code \enzo \ \citep{ENZO_2014}. We point to recent works \citep[e.g. Sec 2.1 in][]{Wittor2019Pol} for the numerical details. \\
 We took three clusters from the \sanpedro-cluster catalogue, which targets the topological study of relics (Wittor et al. in prep). The \sanpedro-simulations use nested grids to provide a uniform resolution at the highest refinement level. Each simulation covers a root-grid of $(140 \ \mathrm{Mpc/h})^3$ and is sampled with $256^3$ cells. Using \MUSIC \ \citep{music}, an additional region of $\sim (4.4 \ \mathrm{Mpc/h})^3-(6.6 \ \mathrm{Mpc/h})^3$ centered around each cluster is further refined using 5 levels, i.e. $2^5$ refinements, of nested grids for a final resolution of $\sim 17.09 \ \mathrm{kpc/h}$. Each nested region is at least $3.5^3$ times larger than the volume enclosed in $r_{200}^3$. We initialised a uniform magnetic field with a value of $10^{-7} \ \mathrm{G}$ in each direction.  \\
 \begin{table}\centering
  \begin{tabular}{lccccc}
   \hline
   ID & $M_{200}$ & $r_{200}$ &  $M_{tracer}$ & $N_p$ & $N_x^3$ \\   
      & $[10^{14} \ \Msun]$ & $[\Mpc]$ &  $[10^{6} \ \Msun]$ & $[10^7]$ &  \\   
   \hline
   SP0m & 5.41 & 1.72 & 3.25 & 1.70 & $256^3$ \\
   SP2m & 8.29 & 1.99 & 5.58 & 2.28 & $320^3$ \\
   SP8m & 7.40 & 1.92 & 3.28 & 1.82 & $384^3$ \\
   E5A & 11.0 & 2.13 & 2.24 & 13.9 & $540^3$ \\
   \hline
  \end{tabular}
  \caption{Overview of the four simulations. ID: cluster name, $M_{200}$: cluster mass,  $r_{200}$: cluster radius, $M_{tracer}$: tracer mass resolution, $N_p$: total number of tracer used, $N_x^3$:  number of resolution elements at the highest nested level of the \enzo-simulation. All values are measured at redshift $z \approx 0$.}
  \label{tab::cluster}
 \end{table} 
 We used cosmological parameters that are based on the latest results from \textit{Planck} \citep[i.e.][]{PlanckVI2018}: $H_0=67.66\ \mathrm{km\ s^{-1}\ Mpc^{-1}}$, $\Omega_{\mathrm{b}}=0.0483$, $\Omega_{\mathrm{m}}=0.3111$, $\Omega_{\mathrm{\Lambda}}=0.6889$, $\sigma_8=0.8102$ and $n = 0.9665$. \\
\begin{figure}
     \centering
     \includegraphics[width = 0.49\textwidth]{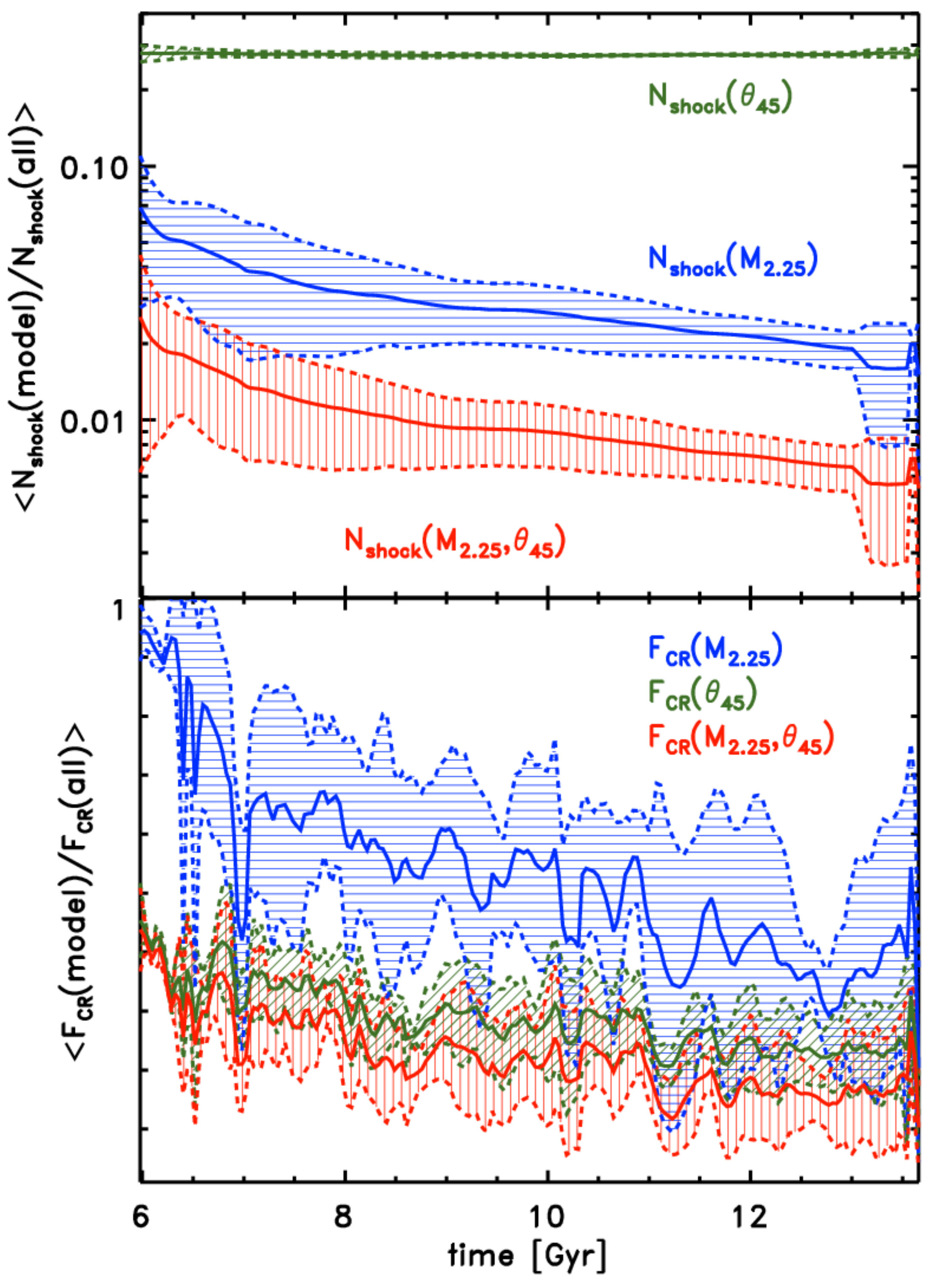}\\
     \caption{Temporal evolution of the shock fraction (top) and Cosmic-ray energy flux ratios (bottom). In both panels, the solid lines show the average of the four simulations and the dashed lines give the standard deviation.}
     \label{fig::flux_ratio}
 \end{figure}
 In this work, we analyse three \sanpedro-clusters, called: SP0m, SP2m and SP8m, that cover a mass range of $\Mth \approx (5.41-8.29) \cdot 10^{14} \ \Msun$ (see Tab. \ref{tab::cluster}). The dynamical state of the three systems is quite active: SP2m and SP8m are both undergoing major merger events close to redshift $z \approx 0$. On the other hand, SP0m hosts a major-merger event at $z \approx 0.4$. 
 \bigskip \\
 We increased our sample by adding one more massive, $M_{200}(z=0) \approx 1.1 \cdot 10^{15} \ \Msun$, cluster (E5A), obtained with earlier \enzo \ simulations and a similar numerical setup  \citep{2018MNRAS.474.1672V,2019arXiv190311052D}. This cluster is an active major-merger and it undergoes various merging episodes including a major merger during its lifetime.  For the further analysis, we use the reconstruction of the sixth AMR level. This assures a uniform resolution of $15 \ \kpc$ across a large enough volume of $\sim(8.5 \ \Mpc)^3$ to follow both components of the major merger. For this work, we used the following cosmological parameters: $h = 0.72$, $\Omega_{\Lambda} = 0.742$, $\Omega_{\mathrm{M}} = 0.258$ and $\Omega_{\mathrm{b}} = 0.0441$.
 \subsection{\CRaTer-simulations}
 We use our Lagrangian code \CRaTer \ to follow the evolution of the shock accelerated cosmic-ray protons in the \enzo-simulations. \CRaTer \ has been already used to study the cosmic rays and turbulence in galaxy clusters \citep[e.g.][]{2016Galax...4...71W,2017MNRAS.464.4448W,2017MNRAS.471.3212W}. For details of the implementation, we point to these references. \\
 At redshift $z \approx 1$, we injected between $\sim (5-28) \cdot 10^6$ particles, with a mass resolution of $\sim (2.24-8.58) \cdot 10^6 \ \Msun$, into each \enzo-simulation. The mass resolution gives a constant and high sampling of the final cluster mass which is crucial to properly model the $\gamma$-emission.  Following the mass inflow, we injected additional tracers from the boundaries at run-time. At redshift $z \approx 0$, each cluster is modelled by $\sim(1.7 - 13.9) \cdot 10^7$ particles. \\ 
 The tracers used a temperature-jump based shock finder to detect shocks in the ICM \citep[Sec. 2.2 in][]{2017MNRAS.464.4448W}. For each detected shock they compute the Mach number:
  \begin{align}
  M = \sqrt{\frac{4}{5} \frac{T_{\mathrm{new}}}{T_{\mathrm{old}}} \frac{\rho_{\mathrm{new}}}{\rho_{\mathrm{old}}} + 0.2}.
 \end{align}
 We defined the corresponding obliquity as the angle between magnetic field and the shock propagation direction which is calculated as the difference between the pre- and post-shock velocities measured by the tracers $\Delta \vvec = \vvec_{\post}-\vvec_{\pre}$:
 \begin{align}
  \theta_{\pre/\para} = \arccos \left( \frac{\Delta \vvec \cdot \bvec_{\pre/\para}}{\left|\Delta \vvec \right| \left| \bvec_{\pre/\para} \right|} \right) .
 \end{align}
 Using the shock velocity and the pre-shock density, we computed the kinetic energy flux across each shock as: $F_{\psi} = 1/2 \cdot \rho_{\pre} v_{\mathrm{sh}}^3$. The associated thermal and cosmic-ray energy flux are: $F_{\mathrm{th}} = \delta(M) F_{\psi}$ and $F_{\mathrm{CR}} = \eta_i(M) F_{\psi}$. The gas thermalization efficiency, $\delta(M)$, was derived by \citet[][]{2013ApJ...764...95K}. For the cosmic-ray acceleration efficiency, $\eta_i(M)$, we assumed a variety of models denoted by $i$. For comparison with \citet{2016Galax...4...71W,2017MNRAS.464.4448W}, we use the efficiencies from \citet{2013ApJ...764...95K}, hereafter $\eta_{13}$. We further tested the new acceleration efficiencies given in \citet{Ryu2019} for the $Q_i = 3.5$ and $p_{\min} = 780 \ \MeV / c$, hereafter $\eta_{19}$, and we combined them with more restrictions on the shock type: Following \citet[][]{2014ApJ...783...91C}, we included the case of only quasi-parallel shocks, i.e. $\theta < 45^{\circ}$, accelerating protons, hereafter $\theta_{45}$. Based on \citet{Ha2018protons}, we only let shocks with $M>2.25$ accelerate cosmic-ray protons, hereafter $M_{2.25}$. Following the definition by \citet{Ha2018protons}, we will refer to shocks with Mach number above $M > 2.25$ as supercritical shocks. Finally, we only let quasi-parallel shocks with $M>2.25$ accelerate cosmic-ray protons, hereafter $\theta_{45},M_{2.25}$. We stress that, unlike \citet{2017MNRAS.464.4448W}, we do not compute effective acceleration efficiencies but use the acceleration efficiencies from the thermal pool. 
 For each different injection model, we computed the time integrated gas and cosmic-ray energy at redshifts $z_j$ as:
 \begin{align}
  E_{\mathrm{gas/CR}} (z_j) = \sum\limits_{z = 1}^{z_j} \sum\limits_{n = 1}^{N_p} F_{i,n}(M, \theta) \Delta t(z) . \label{eq::energies}
 \end{align}
 Depending on the assumed acceleration model, the sum across the particles, i.e. the sum across $n$, is only taken for the shocks that fulfill the conditions summarised.
 \begin{figure*}
 \includegraphics[width = \textwidth]{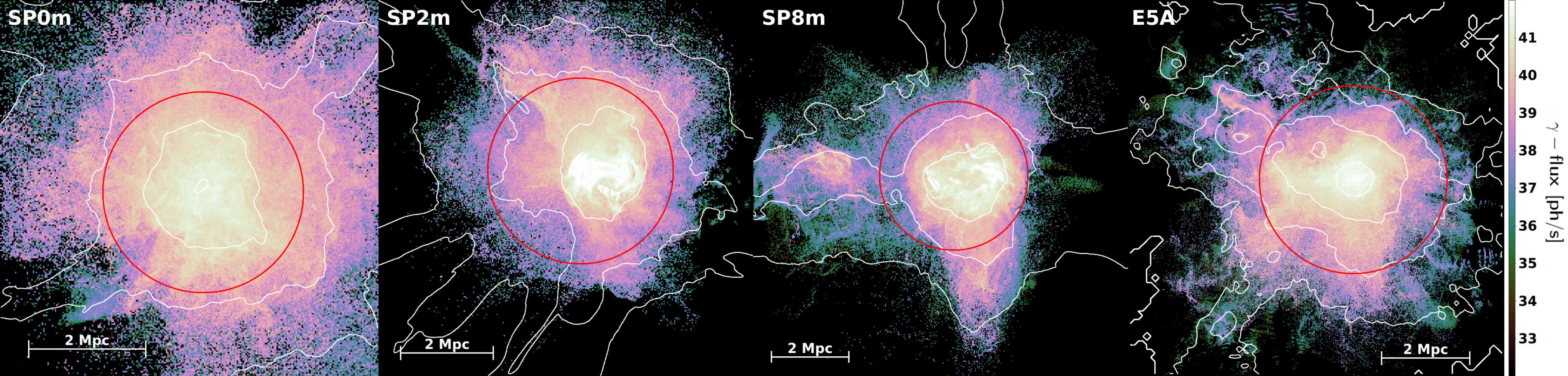}
 \caption{Map of the integrated $\gamma$-ray flux for the $\eta_{19}, \theta_{45}, M_{2.25}$-model. The white contours are the baryonic density at  $[10^{-27.25}, \ 10^{-28}, \ 10^{-28.75} \ \& \ 10^{-29.5}] \ \mathrm{g} / \cm^3$. The red circles mark the $r_{200}$.}
  \label{fig::gammas}
 \end{figure*}
 \section{Results}\label{sec::res}
 \subsection{Shocks \& Cosmic-Ray Energy Flux}
 First, we estimated the number of different types of shocks (e.g. $\theta < 45^{\circ}$, $M>2.25$ etc.) in our simulations. Therefore, we define the shock fraction as the ratio between the number of a specific type of shock and the total number of shocks. In the top panel of Fig. \ref{fig::flux_ratio}, we plot the shock fraction averaged over the four simulations in time. The fraction of quasi-parallel shocks, the green line, stays constant in time, $\sim 28 \ \%$, and also the standard deviation across the different simulations is small. This agrees with  \citet{2017MNRAS.464.4448W}, where we reported that the distribution of shock obliquity follows the distribution of angles between random vectors in a three dimensional space and, hence, about $1/3$ of all shocks are expected to be quasi-parallel. In contrast, the fraction of supercritical shocks decreases in time by a factor of $\sim 4.3$. The fraction of supercritical quasi-parallel shocks follows a similar trend but it drops from $\sim 2.3 \ \%$ to $\sim 0.5 \ \%$. Hence, less than $\sim 3 \ \%$ of all shocks are expected to accelerate cosmic-ray protons.  \\
 To investigate, the amount of the energy processed by shocks in our different models, we define the cosmic-ray energy flux ratio as $F_{\mathrm{CR}}(\mathrm{model})/F_{\mathrm{CR}}(\mathrm{\eta_{19}})$. Here, "model" is replaced by the three models that cut on obliquity, Mach number and both. In the bottom panel of Fig. \ref{fig::flux_ratio}, we show the evolution of the flux ratio averaged over the four simulations. \\
 At early times, most, $\sim (94 \pm 5) \ \%$, of the cosmic-ray energy flux is processed by supercritical shock. This seems obvious, as the cosmic-ray energy flux scales with the shock acceleration efficiency, which increases significantly for supercritical shocks \citep[see Fig. 4 in][]{Ryu2019}. Throughout the cluster evolution, the relative contribution of supercritical shocks shows a large variations with an average value of $\sim 53 \ \%$. Overall, it decreases to $\sim (20 \pm 16) \ \% $ reflecting the occurrence of fewer supercritical shocks at later times. The reduction of supercritical shocks is also shown in the cosmic-ray energy flux processed by quasi-parallel shocks, that is initially $\sim (45 \pm 6) \ \%$ and decreases to $\sim (12 \pm 7) \ \%$. This drop is less than for the supercritical shocks, as high Mach number shocks tend to cluster around structures \citep{2016Galax...4...71W} and, therefore, there can be an excess in supercritical quasi-parallel shocks. On average, the quasi-parallel shocks process $\sim 29 \ \%$ of the cosmic-ray energy flux. This is in agreement with the fact that on average $\sim 27 \ \%$ of the shocks are quasi-parallel. \\
 The cosmic-ray energy flux processed by supercritical quasi-parallel shocks, i.e. shocks that are expected to accelerate cosmic-ray protons, is $\sim (44 \pm 7) \ \%$ at $t\approx 6 \ \Gyr$. The flux ratio follows roughly the same evolution as for the quasi-parallel shocks, though it is on average a factor of $\sim 1.3$ smaller, and it drops to $\sim (10 \pm 7) \ \%$ at the end of the simulations. The amount of flux processed by supercritical quasi-parallel shocks is larger than what one would expect from the average number of these kind of shocks. Again, this reflects the clustering of supercritical shocks around structures. 
 \subsection{$\gamma-$ray emission}
\begin{figure}
  \includegraphics[width = 0.49\textwidth]{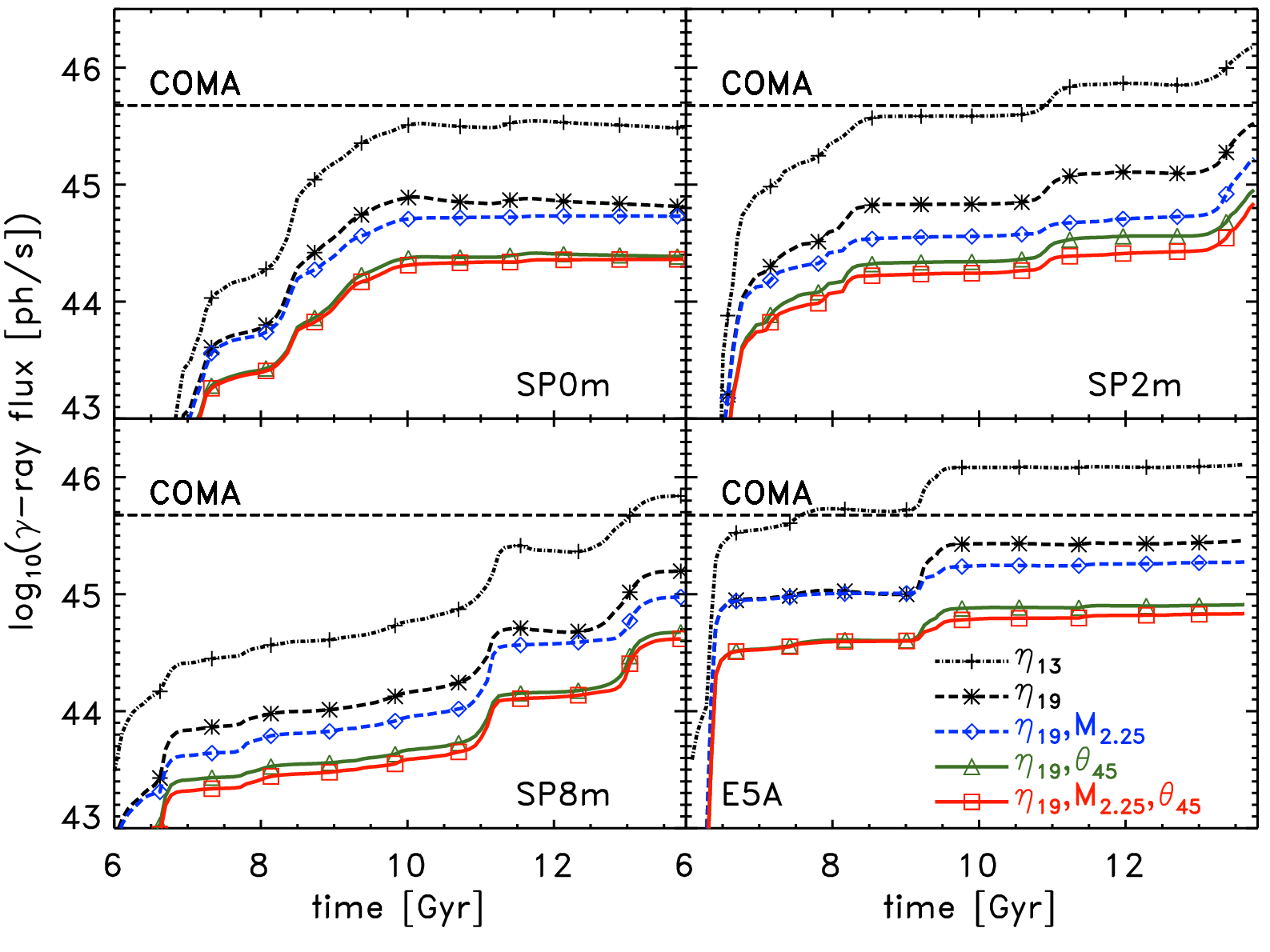}
    \caption{Time evolution of the $\gamma$-ray flux for the different cosmic-ray models and the COMA upper limit \citep[dashed line][]{2016ApJ819149A}.}
  \label{fig::gamma_flux}
 \end{figure}

\begin{figure}
    \includegraphics[width = 0.49\textwidth]{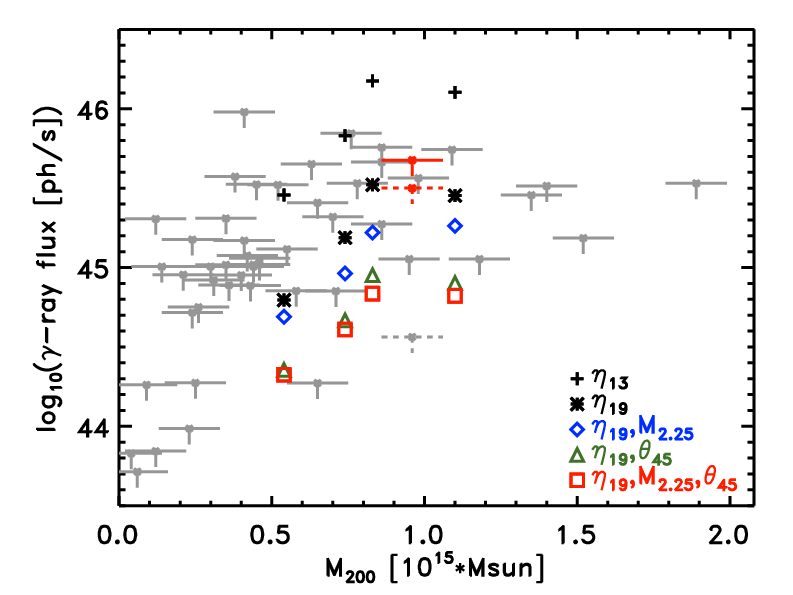}
  \caption{ The total $\gamma$-ray flux within $r_{200}$ vs the cluster mass $M_{200}$. The grey crosses give the upper limits in the energy range $500 \ \MeV  -  200 \ \GeV$ \citep{2014ApJ78718A} . Specifically, the grey dashed cross is the upper limit for the COMA. The red solid cross gives the deeper limits for COMA \citep{2016ApJ819149A} and the red dashed cross is an estimate for the improvement on the upper limit after ten more years of Fermi observations. The symbols give the $\gamma$-ray flux of the different models in the simulations.}
  \label{fig::gamma_limits}
 \end{figure}

 We computed the $\gamma$-ray emission for each cosmic-ray population as described in Sec. C1 in \citet{2017MNRAS.464.4448W} and references therein \citep{2013A&A...560A..64H,2015MNRAS.451.2198V,2016MNRAS.462.2014D}. Given the large cooling timescales of cosmic-ray protons, we did not include any energy losses and our results impose upper limits on the $\gamma$-ray flux. We based our energy range, $100 \ \MeV$ to $10 \ \GeV$ using $16$ energy bins, similar to Fermi \citep{2016ApJ819149A}. In Fig. \ref{fig::gammas}, we give maps of the $\gamma$-ray emission for the most restricted cosmic-ray model, i.e. $\eta_{19},\theta_{45},M_{2.25}$. In addition, we show the baryonic densities contours and the region of $r_{200}$. In all cases, $\gamma$-ray emission is mostly confined within $r_{200}$. Moreover, the peak of $\gamma$-ray emission always overlaps with the densest regions. In E5A, the $\gamma$-ray emission is more prominent in the right sub-cluster.\\
 In Fig. \ref{fig::gamma_flux}, we plot the evolution of the $\gamma$-ray emission for each cosmic-ray model measured in the whole simulation box, as well as the upper limit on the COMA cluster given in \citet{2016ApJ819149A}. For all clusters except SP0m, which is much lighter than COMA, the $\eta_{13}$ model produces $\gamma$-ray emission that is above the COMA limit. In all clusters, the $\gamma$-ray emission associated with the $\eta_{19}$ model about a factor of $\sim 4.5$ smaller than in the $\eta_{13}$ model and, hence, it drops in always below the Fermi-limit of COMA. Additional cuts in Mach number and/or obliquity reduce the $\gamma$-ray emission further by factors of $\sim 6.5 - 23.4$ compared to the $\eta_{13}$ model. \\
 This suggests that already the $\eta_{19}$-model might explain the non-detection of $\gamma$-rays. In order to exam if the simulated clusters could be observed by Fermi, we compared our sample with \citet{2014ApJ78718A} who set upper limits on the $\gamma$-ray flux in the energy range $500 \ \MeV - 200 \ \GeV$, measured in $r_{200}$, for a larger sample of clusters. In Fig. \ref{fig::gamma_limits}, we plot the cluster mass against the upper limit\footnote{We used the $M_{200}$ values from Tab. 1 and $F^{UL}_{\gamma,500 \ \MeV}$ values of an extended source from Tab. 6 both given in \citet{2014ApJ78718A}.}. In addition, we plot the upper limit for the COMA cluster as given in \citet{2016ApJ819149A}\footnote{We refer to the value given in Tab. 1. for a cored profile with $\Gamma = 2.3$ which is similar as in our simulation.} and an estimate for the COMA upper limit after ten years of Fermi observations\footnote{We re-scaled the upper limit from \citet{2016ApJ819149A} given the estimates in Fig. 2.15.1 in \citet{eastrogram}.}. We note that the deeper upper-limit for COMA is larger than the limit in \citet{2014ApJ78718A}, as \citet{2016ApJ819149A} are probing lower energies that have more low-energy particles for a power-law distribution. For a proper comparison with the limits given in \citet{2014ApJ78718A}, we additionally computed the $\gamma$-ray flux in the same energy range. Yet, the $\gamma$-ray fluxes do not change significantly, as the effective cross-section \citep[Eq. (79) in][]{2006PhRvD..74c4018K} peaks at $1.22 \ \GeV$ and rapidly decreases for lower energies. This has been also observed in other works \citep[e.g.][]{Brunetti_Zimmer_Zandanel_2017}. Hence, in Fig. \ref{fig::gamma_limits}, we plot the, somewhat larger, $\gamma$-ray fluxes in the energy energy range of $100 \ \MeV - 10 \ \GeV$ measured inside $r_{200}$, i.e. only inside the red circle in Fig. \ref{fig::gammas}. \\
 If our simulated clusters are compared to the available mass-$\gamma$-flux relation derived from Fermi observations, all of them should be detected by Fermi for the $\eta_{19}$-model (black asterisks). If only supercritical shocks accelerate protons (blue diamonds), the $\gamma$-ray flux is reduced by a factor of $\sim 2$ at the most and all clusters remain detectable. The additional cut in obliquity (red squares) lowers the $\gamma$-ray flux significantly and, without exceptions, the simulated clusters drop below the Fermi-limits. Only SP0m remains at the edge of what could be observable, as its $\gamma$-ray flux is above the upper limit of Abell 2877. Though, we found fairly little information about Abell 2877 in the literature. Hence, it is not clear if Abell 2877 and SP0m are in a similar dynamical state and if their evolution is comparable. Hence, one should not take this data point too strict. If only quasi-parallel shocks of all strengths accelerate protons (green triangles), the $\gamma$-ray flux is slightly above the $\gamma$-ray flux of the $\eta_{19},M_{2.25},\theta_{45}$-model. Hence, the $\gamma$-ray emission is mostly reduced by the cut in obliquity, which is already enough to explain the non-detection of $\gamma$-rays. \\
 As an alternative approach, also the stacking of $\gamma$-ray count maps has been explored to produce statistical upper limits for the average cluster population \citep{2012A&A...547A.102H,2013A&A...560A..64H}. We performed a qualitative assessment of stacking on our clusters, by computing the arithmetic mean of the $\gamma$-ray emission in a region of size $\sim (2.13 \ \Mpc)^2$ centered around our simulated clusters. We find a mean flux is $\sim 4.16 \cdot 10^{44} \ \ph / \sek$, which is still below the deep COMA limit of \citet{2016ApJ819149A}. The stacked signal of the simulations is also below the upper limit of \citet{2013A&A...560A..64H}, when re-scaled to the same energy range by assuming a spectral index of $2.3$. \\
 In summary, although our sample is still fairly limited to fully compare with the Fermi sample, the observed trends are regular enough to derive some fiducial conclusions on the tested DSA models. We find that the acceleration efficiencies presented in \citet{Ryu2019} can explain the non-detection of $\gamma$-rays, if only quasi-parallel shocks with Mach numbers above $M>2.25$ are able to accelerate cosmic-ray protons. Furthermore, Fermi would not be able to detect any of our simulated clusters, if it observers for ten more years (red dashed cross). 
 \section{Conclusion}\label{sec::conc}
 In this work, we have re-investigated the puzzle of the missing hadronic $\gamma$-ray emission in galaxy clusters \citep[e.g.][]{2016ApJ819149A}. We have simulated evolution of cosmic-ray protons accelerated according to different possible recipes of DSA in  four galaxy clusters, using \enzo \  and our Lagrangian tracer code \CRaTer. We followed each cluster simulation using about $10^7-10^8$ tracer particles that recorded shocks and computed the associated cosmic-ray energy injected by the shock. We then computed the corresponding $\gamma$-ray flux for the different models for the shock acceleration efficiencies and compared this to the upper limits set by Fermi. Our findings are summarised as follows:
 \begin{itemize}
     \item The $\gamma$-ray flux within $r_{200}$ produced with the new acceleration efficiencies, that account for the dynamical feedback of cosmic-ray pressure on the shock, is reduced a factor of $\sim 4.5$ smaller on average; such reduction is however not enough to make all our simulated clusters invisible to Fermi.
    \item As most of the cosmic-ray energy flux is processed by shocks with $M>2.25$, the $\gamma$-ray emission is not reduced significantly and the clusters remain detectable in $\gamma$-rays, if only $M<2.25$ shocks are excluded from cosmic-ray protons acceleration.
    \item While allowing cosmic-ray proton acceleration only at quasi-parallel shocks produces $\gamma$-ray emission close to but above the Fermi-limits, the predicted $\gamma$-flux drops below the Fermi-limits if acceleration is additionally restricted to supercritical shocks only.
 \end{itemize}
 In conclusion, we have found that the new acceleration efficiencies given in \citet{Ryu2019} might explain the non-detection of $\gamma$-rays in galaxy clusters, but only if supercritical quasi-parallel shocks, i.e. $M>2.25$ and $\theta < 45^{\circ}$, are able to accelerate cosmic-ray protons. Besides, the new acceleration efficiencies, the main contributor for reducing the $\gamma$-rays flux is the cut in obliquity, as most of the energy is being processed by $M>2.25$ shocks. Though, only the detection of a $\gamma$-ray signal can verify the exact acceleration scenario of cosmic-ray protons. Our estimates show that even ten more years of observation with Fermi are most-likely not enough to make any significant detection and one would require more sensitive telescopes. As the CTA will be most-likely less sensitive than Fermi in the desired energy range \citep{CTA}, future hopes to detect $\gamma$-rays in galaxy clusters are restricted to the proposed mission AMEGO \citep{amego}.\\
 As final caveats, we note that we did not include $\gamma$-ray emission associated with the turbulent (re-)acceleration of cosmic-ray protons \citep[e.g.][]{Brunetti_Zimmer_Zandanel_2017} which could increase the observed signal. Furthermore, we did not include possible re-acceleration of cosmic-ray protons at shocks, either. Assuming that it operates only at supercritical quasi-parallel shocks, \citet{Ha2019gammas} estimated that re-acceleration would increase the cosmic-ray energy by $\sim (40-80) \ \%$. In general, it is argued that, at quasi-perpendicular shocks, protons go about one gyromotion in the shock foot and then advect downstream. Yet at subcritical shocks, overshoot/undershoot oscillations do not developed in the shock transition and the specular reflection of protons is negligible. Hence, upstream waves are not generated there. However, it remains unknown if pre-existing cosmic-ray protons can be reflected efficiently at either quasi-perpendicular or subcritical shocks, leading to the self-excitation of upstream waves. Hence, the re-acceleration by turbulence and at shocks need further investigations.
 Finally, we comment that the numerical resolution appears sufficient for the presented analysis as the average cosmic-ray dynamic is likely to be converged, and that increasing the resolution would only affect the cluster cores, based on previous studies \citep[e.g.][]{2014MNRAS.439.2662V}.
\section*{acknowledgments}
  The authors gratefully acknowledge the Gauss Centre for Supercomputing e.V. (www.gauss-centre.eu) for supporting this project by providing computing time through the John von Neumann Institute for Computing (NIC) on the GCS Supercomputer JUWELS at J\"ulich Supercomputing Centre (JSC), under projects no. 11823, 10755 and 9016 as well as hhh42,  hhh44 and stressicm. \\ 
  D. W. and  F.V. acknowledge financial support from the European Union's Horizon 2020 program under the ERC Starting Grant "MAGCOW", no. 714196. We also acknowledge the usage of online storage tools kindly provided by the Inaf Astronomica Archive (IA2) initiave (http://www.ia2.inaf.it). H.K. was supported by the Basic Science Research Program of the National Research Foundation of Korea (NRF) through grant 2017R1D1A1A09000567. D.R. was supported by the NRF through grants 2016R1A5A1013277 and 2017R1A2A1A05071429. We thank Filippo D’ammando for useful insights on $\gamma$-ray telescopes. We gratefully acknowledge the {\enzo} development group for providing online documentations. 
 \bibliographystyle{mnras}
 \bibliography{mybib}
 
\end{document}